\documentclass[aip,apl,preprint]{revtex4-1}

\usepackage{xcolor}

\usepackage{graphicx}
\usepackage{dcolumn}
\usepackage{amsmath}
\usepackage{bm}

\usepackage{ulem}

\def\beq{\begin{eqnarray}}
		\def\eeq{\end{eqnarray}}
	\def\be{\begin{equation}}
		\def\bel{\begin{equation}\label}
			\def\beel{\begin{eqnarray}\label}
				\def\ee{\end{equation}}

			\def\bm{\begin{math}}
				\def\me{\end{math}}

            \def \e{\varepsilon}

			\def\D{\Delta}
			
			\newcommand \nn {\nonumber}
			\newcommand \bei {\begin{itemize}}
				\newcommand \eei  {\end{itemize}}

\begin{document}
            

\title{Investigation of filamentation in a-Si/Ag/Cu memristors with atomic force microscope} 



\author{A. Samsonova}
    \email[]{alena.samsonova@skoltech.ru}
    \affiliation{Skolkovo Institute of Science and Technology, Bolshoy Boulevard 30, bld. 1, Moscow, Russia 121205}
\author{V. Dremov}
    \affiliation{Moscow Institute of Physics and Technology, Institutskiy pereulok 9, Dolgoprudny, Russia 141700}
\author{O. Klimenko}
    \affiliation{Skolkovo Institute of Science and Technology, Bolshoy Boulevard 30, bld. 1, Moscow, Russia 121205}
    \affiliation{PN Lebedev Physical Institute of RAS, Leninskiy prospekt 53, Moscow, Russia 119991}
\author{N. Brilliantov}
    \affiliation{Skolkovo Institute of Science and Technology, Bolshoy Boulevard 30, bld. 1, Moscow, Russia 121205}
\author{V.N. Antonov}
    \affiliation{Skolkovo Institute of Science and Technology, Bolshoy Boulevard 30, bld. 1, Moscow, Russia 121205}
    \affiliation{Moscow Institute of Physics and Technology, Institutskiy pereulok 9, Dolgoprudny, Russia 141700}
    


\date{4 May 2026}

\begin{abstract}
Cation-based  Ag/Cu filaments formed in an insulating $\alpha$-Si matrix are widely used as memristors in crossbar arrays for efficient in-memory computing. However, the stochastic nature of filament formation and rupture gives rise to device-to-device and cycle-to-cycle variation. Despite successful implementation of large-scale memristor arrays, systematic studies of filament parameters and their spatial distribution in the memristors are scarce. In this work, we use conductive atomic force microscopy (c-AFM) to probe the spatial distribution of conductive filaments in $\alpha$-Si memristors. The charge transport is dominated by a limited number of discrete filaments rather than by uniform conduction across the device area. The systematic analysis of the experiment gives the mean surface density of the conductive filaments $\sim$3200 per $\mu\text{m}^2$. Both volatile and non-volatile filaments can be found within one memristor. The experimental data and the nature of volatile and non-volatile filaments may be explained within the model of multiple trap assisted tunnelling. The model yields reasonable estimates for physical properties for both types of filaments.
\end{abstract}

\pacs{}

\maketitle 


Memristors have been extensively studied as elements for a next-generation computing system that would overcome the limitations of traditional von Neumann architecture due to their advantages such as nonvolatility, low energy consumption, high integration density, and CMOS compatibility~\cite{Ielmini2018, Xia2019, Xiao2023, Aguirre2024}. Crossbar arrays of memristors are particularly well-suited for implementing neural networks, as they can efficiently perform in-memory vector-matrix multiplication and mimic neuronal functions~\cite{Wang2019, Ielmini2021, Leng2022}. However, the main roadblock towards further development and commercialisation of memristor crossbar arrays is the sneak-path currents that limit the large-scale integration of memristors and the stochasticity of filament formation. The memristors have large cycle-to-cycle and device-to-device variation~\cite{Yang2014, delValle2018, Zhang2020}. While the problem of sneak-path currents can be solved with the use of a transistor, selector, or self-rectifying memristor~\cite{Wang2019_b, Zhang2024, Samsonova2025}, overcoming stochasticity requires engineering memristors with predictable switching parameters, which is challenging. Currently, the understanding of filament formation and ion migration, which are precursors to resistive switching, is poor. \par 
Resistive switching in cation-based memristors is caused by the formation of conducting filaments composed of spatially separated nanoparticles of electrochemically active metals, such as Ag or Cu~\cite{Jo2009, Valov2011, Liu2012, Chae2017}. The factors influencing this process are the kinetic parameters: ion mobility and redox rate~\cite{Yang2014}. In switching media with low ion mobility, such as amorphous Si ($\alpha$-Si), Ag cations are reduced within the dielectric, leading to the growth of conducting filaments from the active electrode to the inert electrode in the form of metal clusters~\cite{Yang2012, Tian2014, Wang2016}. The stability of Ag conductive filaments in $\alpha$-Si can be fine-tuned by incorporating transition metal elements that form stable silicides with Si~\cite{Lubben2020, Yeon2020, Kang2022}. In such a case, transition-metal clusters act as backbones for the conductive filaments within the dielectric, thereby enhancing the uniformity and stability of resistive switching. \par 

In situ observation techniques, such as TEM, STM, or conductive AFM, together with the theoretical models, would provide valuable information on the nature of resistive switching~\cite{Yang2018}. Among these, conductive AFM enables determining the current distribution across the memristor area, evaluating the density of the conductive filaments, and identifying their locations~\cite{Lanza2017, Ang2019, Qi2020}. For example, conductive AFM reveals oxygen ion migration and the formation of oxygen-deficient filaments in HfO$_2$-based memristors~\cite{Yang2017}. \par 

In this work, we determine whether conduction is confined to a number of individual filaments or occurs uniformly across the entire area of the memristor. Using conductive AFM, we visualise the filament distribution and carry out a statistical analysis of their electrical properties. Ultimately, by confining filament formation to the localised regions, we can engineer memristors with improved uniformity and stability of resistive switching.

 
\begin{figure}[b!]
    \centering
    \includegraphics[width=\textwidth]{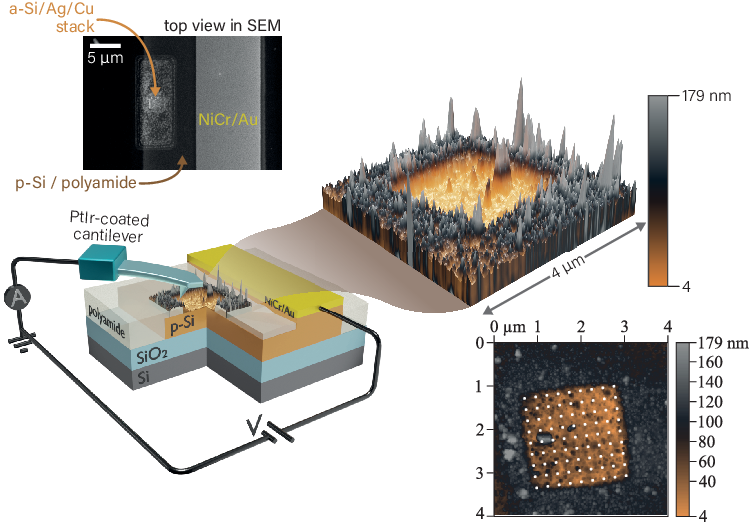}
    \caption{Cross-section and AFM topography (4$\times$4~$\mu$m$^2$) of the $\alpha$-Si-based memristor on a SOI wafer with a highly boron-doped p-Si. The PtIr-coated conductive cantilever is used for current mapping in contact mode. The SEM image in the top-left shows a smooth and continuous NiCr/Au thin film and a granular morphology of the $\alpha$-Si/Ag/Cu stack. Figure at the bottom right shows an AFM topography image with a position of $9\times9$ grid of measured points (spacing $\approx$ 250~nm and cantilever-sample contact area $=29\pm6$~nm$^2$)}
    \label{fig1}
\end{figure}

Our standard fabrication route for fully operational $\alpha$-Si-based memristors on a SOI wafer with a highly boron-doped top Si begins with etching the p-Si to obtain the desired thickness of 1~$\mu$m. This p-Si is then used to form the bottom electrode of the memristor using UV lithography; see the bottom left of Fig.~\ref{fig1}. Subsequently, the NiCr/Au (10~nm/90~nm) contact pads are deposited via thermal e-gun evaporation on top of p-Si. Next, a polyimide is spin-coated over the entire wafer to electrically isolate between the bottom electrode from the next layers. Window $2\times2$~$\mu$m$^2$ is opened in the polyimide next to NiCr/Au metallization with plasma etching, where the $\alpha$-Si/Ag/Cu (6~nm/3~nm/2~nm) stack is formed by AC/DC magnetron sputtering of metals. Finally, a set of top Al cross electrodes is typically patterned with UV lithography and deposited via thermal e-gun evaporation. However, in this work, we stop processing at the previous step to allow direct access to the $\alpha$-Si/Ag/Cu stack with conductive AFM (c-AFM measurements). The cross section of the device, the AFM topography and SEM images of the memristor stack are shown in Fig.~\ref{fig1}. The SEM image shows the smooth surface of the NiCr/Au electrode, in contrast to the granular nature of the $\alpha$-Si/Ag/Cu stack. 

AFM acquires surface topography in tapping mode and the current mapping in contact mode using c-AFM functionality. For c-AFM measurements, we use PtIr-coated conductive probes with a tip curvature radius of 35~nm and a typical spring constant of 11.8~N/m. The current is measured over the $\alpha$-Si/Ag/Cu stack in the polyimide window. A bias voltage, ranging from -10~V to +10~V, is applied to the NiCr/Au electrode of the memristor, while the conductive tip is grounded, as shown in Fig.~\ref{fig1}. The current amplification is limited with the compliance of 500~nA. \textit{I-V} curves are measured by the voltage sweeps from positive to negative values and backwards. It takes several seconds to take each \textit{I-V} curve. All scans are carried out under the ambient conditions.


\begin{figure}[t!]
    \centering
    \includegraphics[width=\textwidth]{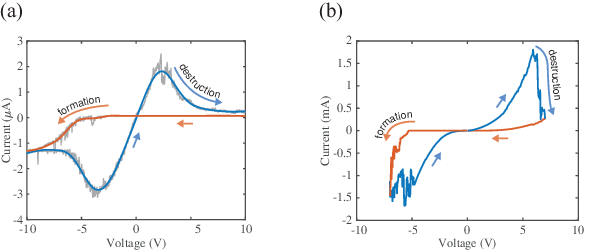}
    \caption[]{(a) Cumulative \textit{I-V} curve obtained by summing all \textit{I-V} curves acquired at different points; (b) The typical \textit{I-V} curve of a fully operational $\alpha$-Si/Ag/Cu memristor with top Al cross electrode~\cite{Samsonova2024}}
    \label{fig2}
\end{figure}

Figure~\ref{fig1} shows the AFM topography of the $\alpha$-Si/Ag/Cu stack. It has a granular morphology with a mean surface roughness $S_a$ of 27$\pm$8~nm. This roughness is attributed to the etching of p-Si prior to the bottom electrode patterning. To investigate the electrical properties of the memristor stack, we defined a $9\times9$ grid of points across the device area, with a step of about 250 nm, as shown in Fig.~\ref{fig1}. \textit{I-V} curves are measured at each grid point using the contact mode of AFM. The curves can be classified into the four categories: (1) non-volatile with well-defined switching events; (2) volatile with no stable resistance state at low voltage; (3) electroforming-like with unstable current spikes; (4) insulator-like with negligible current even at high bias. The cumulative current summed across all grid points results in a non-volatile curve shown in Fig.~\ref{fig2}a. Although it is consistent with fully operational $\alpha$-Si/Ag/Cu memristors shown in Fig.~\ref{fig2}b~\cite{Samsonova2024}, the current is lower by $10^3$. This difference will be discussed below. \par 

\begin{figure}[t!]
    \centering
    \includegraphics[width=\textwidth]{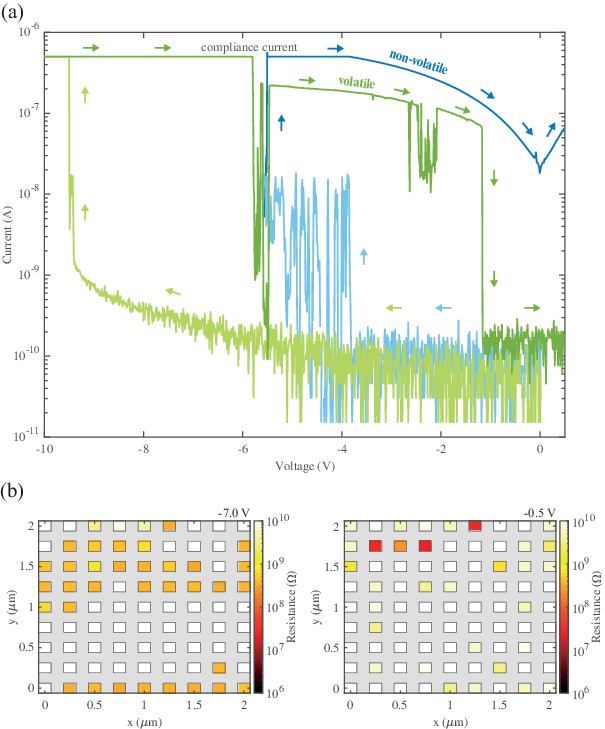}
    \caption[]{(a) Typical \textit{I-V} curves of non-volatile (blue) and volatile (green) filaments; (b) Resistance maps at -7.0~V and -0.5~V show the distribution of conductive filaments. Measurements are taken on a 9$\times$9 grid with 250 nm spacing between the points. The estimated contact area is 29$\pm$6~nm$^2$}
    \label{fig3}
\end{figure}

Figure~\ref{fig3}a shows typical volatile and non-volatile \textit{I-V} curves  measured at two different grid points. Both exhibit a switching-on event, but only the non-volatile curve retains the conductive state after the bias voltage removal. Both volatile and non-volatile filaments contribute to resistive switching, which are the two limiting cases of the filamentary mechanism. While non-volatile filaments are stable and retain their properties after the bias removal, volatile filaments remain conductive solely at a high bias voltage. Across 7 studied samples, only a small fraction of pixels, about 9\%, exhibit non-volatile switching, attributed to the formation and rupture of conductive filaments. The mean set voltage for the 52 non-volatile pixels is -5.2$\pm$2.4~V, determined at a current threshold of 10~nA. There are also 6 extra pixels that are initially formed, $R \sim$ tens of M$\Omega$. At -0.5~V, the mean resistance of non-volatile filaments is about 41~M$\Omega$. In contrast, the fraction of pixels with volatile filaments is dominant, reaching 38\%, with a mean set voltage of -6.2$\pm$2.4~V. In some areas, there is an electroforming-like behaviour or breakdown spikes (22\%), while up to 30\% of the pixels remain non-conductive. The ratio of pixels with different behaviour varies between the samples, reflecting the stochasticity of the filament formation. \par

We construct resistance maps to visualise the spatial distribution of the conductive filaments by extracting resistance at two bias voltages, -7.0~V and -0.5~V, in the reverse branch; see Fig.~\ref{fig3}b. The maps highlight the spatial distribution of filaments and the variation of resistance. Although many pixels exhibit low resistance at high negative voltage, indicating the formation of conductive filaments, the majority of them are volatile. They dramatically lose their conductance at low bias, making the spatial distribution of filaments across the stack even more inhomogeneous. \par 

For later analysis we estimate the cantilever-sample contact radius $a_{JKR}$ using the Johnson-Kendall-Roberts (JKR) model~\cite{JKR:1971}:
\begin{equation}
    a_{JKR}=\left[ \frac{3R_{eff}}{4E^\star} \left( F_N + 2F_{ad}+\sqrt{4F_NF_{ad}+4F_{ad}^2}\right) \right]^{1/3},
\end{equation}
where $R_{eff}=\left[ 1/R_1+1/R_2\right]^{-1}$ is the effective curvature radius, with $R_1$ and $R_2$ standing for the cantilever tip and sample curvature radii. $E^\star=\left[ (1-\nu_1^2)/E_1+(1-\nu_2^2)/E_2 \right]^{-1}$ is the effective modulus, where $E_{1}=168~\text{GPa}$ and $\nu_{1}=0.39$ are Young's modulus and Poisson's ratio of the cantilever tip (Pt), and $E_{2}=100~\text{GPa}$ and $\nu_{2}=0.49$ are those of the sample (Cu). $F_N$ is the applied normal force and $F_{ad}$ is the pull-off (adhesion) force of the tip-sample interaction. Assuming a flat sample surface, we take $R\rightarrow R_{tip}=35~\text{nm}$, which is the tip curvature radius. The adhesion force of the Pt/Cu contact is $12\pm3~\text{nN}$, and the normal force varies within the range of 10--20~nN. The contact radius yields a contact area of $s_0=29\pm6~\text{nm}^2$. \par

To plot Fig.~\ref{fig2}a we added $I-V$ curves of 15 non-volatile pixels and 54 volatile pixels found in one of the samples (we apply binary mapping with a threshold resistance of $50~\text{M}\Omega$ to identify the number of non-volatile filaments), which resulted in a total current of a few $\mu$A. At the same time, the current in $I-V$ curve of fully operating memristor is three orders larger, Fig.~\ref{fig2}b. To explain the difference, we assume that the density of the conducting filament is approximately uniform across the memristor area. The total contact area, associated with 81 pixels, is estimated to be $81\, s_0=81\times29\, \text{nm}^2=2.349\cdot 10^3\,\text{nm}^2$, while the total area of the memristor is $2\times2~\mu\text{m}^2=4\cdot10^6\,\text{nm}^2$, which is three orders of magnitude larger. Hence, the current in Fig.~\ref{fig2}a would also be three orders of magnitude larger, in agreement with the observation data in Fig.~\ref{fig2}b.

To gain a deeper understanding of the nature of conducting filaments, we consider electron transport there and assume the trap assisted tunnelling (TAT) model~\cite{SCHENK19921585, Nasyrov2003, IELMINI20021749, Luca}. In this model, the conduction through a filament occurs via  electron tunnelling from an active electrode to the closest trap, and then via successive tunnelling between the traps. As we show below, the largest contribution to the transport is made by the tunnelling between the bottom (inactive) electrode and the trap, closest to this electrode. So far TAT theories have been developed for traps in a crystalline dielectric with well-formed bands (valence and conduction)~\cite{SCHENK19921585, Nasyrov2003, IELMINI20021749, Luca}. The emission and capture of electrons by traps occurs in such a system through the conduction band. In the amorphous dielectric, such a mechanism is absent, so one needs to revise the theory. Here we present a qualitative analysis of TAT in the amorphous dielectric. A comprehensive theory will be published elsewhere.

Since electrons lose coherence when hopping between traps, one can  use the Landauer formula~\cite{nazarov_blanter_2009} for the conductance of each segment between the $i$-th and $i+1$th trap, $G_{i,i+1}=N_{\rm open}^iT_{i,i+1}e^2/h$, where $e^2/h$ is the conductance quanta,  $N_{\rm open}^i$ is the number of open conduction channels for this segment, and $T_{i,i+1}$ is the respective transmission coefficient, which we assume to be equal for all open channels. Further, by assuming a thermal equilibration of electrons between hopping events, one can apply the resistance summation, yielding the resistance of a filament, comprised of $M$ traps (we estimate $M\sim$ 10-12), 
\beel{eq:Res1}
R_{\rm fil}&=& G^{-1}_{el, 1}+ G^{-1}_{M,el} +\sum_{i=1}^{M-1}G_{i,i+ 1} \\
&=&
\frac{h}{e^2N_{\rm open}^{1} T_{el,1}}+
\frac{h}{e^2N_{\rm open}^{M} T_{M,el}} + \sum_{i=1}^{M-1} \frac{h}{e^2N_{\rm open}^{i} T_{i,i+1}}, \nn
\eeq
where the first two terms in the r.h.s. of (\ref{eq:Res1}) refer to the conducting segments between the active electrode and the first trap and the last $M$th trap and the second (bottom) electrode; $N^i_{open}$ is the number of open conductance channels of the filament for the $i$th segment. \par
The distance between the successive traps $\D l(x)$ increases drastically with increasing distance from the active electrode $x$, approximately as $\D l(x) \sim \exp (a x^2)$, where $a$ is some constant.  Indeed, the mean distance between the traps is expressed as $\sim n(x)^{-1/3}$, where $n(x)$ is the respective  trap density, while the diffusional profile of the traps scales as $n(x)\sim \exp(- x^2/4Dt) $, where $D $ is the ions' diffusion coefficient (recall that the filaments are formed by the ions' diffusion). The tunnelling conductance, in turn, decreases exponentially with $\D l(x)$: $T\sim \exp(-b \D l(x))$ ($b$ is a constant).  Hence, we expect that the conductance between successive traps, located  at the distance $x$ from the active electrode, scales as $G_{i,i+1}(x) \sim \exp(-b e^{ax^2})$. In practice, this implies that the total conductance of a filament is limited either by the inter-trap tunnelling between the  last two traps ($M-1$th and $M$th), or between the last trap and the bottom electrode. Without loss of accuracy, we assume the latter: the conductance is limited by the tunnelling between the last trap and the bottom electrode. Then the first term and the sum in (\ref{eq:Res1}) may be neglected:
\bel{Res}
R_{\rm fil} \approx 
\frac{h}{e^2N_{\rm open}^{M} T_{M,el}} = \frac{h}{e^2N_{\rm open} T}, 
\ee
where we use shorter notation. 

The number of open channels $N_{\rm open}$ in each filament can be estimated from the extension of the electron wave-function, $\psi(r) \sim r^{-1} e^{-r/r_0}$, localised on the trap~\cite{Lundsr1972, SCHENK19921585}, associated with Ag$^+$ ion (or, possibly, with ion's cluster), that is, the diameter of each filament is $d=2r_0$~\cite{Lundsr1972, SCHENK19921585}:

\bel{5}
r_0^{-1}=\frac{\sqrt{2m_*(U_0- \varepsilon)}}{\hbar}.
\ee
Here $\varepsilon $ is the energy of the particle in the trap,  $U_0$ is the height of the potential barrier, associated with the dielectric  between the metal electrodes, and $m^*\approx 0.3 m_e$ is the effective mass of electron in $\alpha$-Si.
The number of open conduction channels in the filament reads~\cite{nazarov_blanter_2009}:
\begin{equation}
\label{Nopen}
N_{\rm open}\sim (k_F \cdot d /\pi)^2 \sim \frac{2\hbar^2 k_F^2}{ \pi^2 m_*(U_0-\varepsilon)}
\end{equation}
In (\ref{Nopen}) $k_F$ is the Fermi wave-vector of the  electrode, taken here for Ag (typically $\sim12~\text{nm}^{-1}$). The transmission coefficient, dependent on the applied voltage $V$ and the distance $l$ between the last $M$th trap and the bottom electrode, may be calculated within the WKB approximation. Taking the electric field as $E=V/d_{Si}$ ($d_{Si}\simeq 6~\text{nm}$ is the thickness of $\alpha$-Si layer), the transmission coefficient, subject to the pre-exponential factor, reads 

\begin{figure}[t!]
   \centering
   \includegraphics[width=0.5\textwidth]{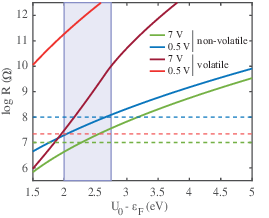}
    \caption[]{The dependence of the filament resistance on the energy gap $U_0-\varepsilon_F$ for non-volatile filaments with the length gap of $l_{nonvol}=1.1\,\text{nm}$ and volatile filaments with $l_{vol}=2.3\,\text{nm}$. The dashed horizontal lines are guidelines for average filaments' resistance for the bias voltage of $V=7\,\text{V}$ (accuracy is limited by 500~nA compliance current set by an experimental setup) and $V=0.5\,\text{V}$. The shaded region, $2.00-2.75\,\text{eV}$, corresponds to the range of gap energies consistent with the experimental data}
    \label{fig4}
\end{figure}

\begin{equation}
\label{R2}
T(V,l) = \exp \left\{- \frac{4 d_{Si} \sqrt{2m^*}}
{3\hbar V} \left[(U_0-\varepsilon)^{3/2} - q\cdot(U_0-\varepsilon - lV/d_{Si} )^{3/2}\right]\right\},  
\end{equation}
where the factor $q=\Theta(U_0 -  \varepsilon-lV/d_{Si})$, with $\Theta(x)$ being the Heaviside step-function, accounts for both forms of the barrier, a triangle and a trapezoidal one. In \eqref{R2} one can use the Fermi energy, $\varepsilon_F$,  in Ag  for $\varepsilon$. Finally, we obtain for the resistance of a single filament: 
\bel{Res2}
R_{\rm fil}(l)=\frac{\pi^3 m_* (U_0-\varepsilon_F)}{e^2 \hbar k_F^2}  \exp \left\{ \frac{4 d_{Si} \sqrt{2m^*}}
{3\hbar V} \left[(U_0-\varepsilon_F)^{3/2} - q\cdot(U_0-\varepsilon_F - lV/d_{Si} )^{3/2}\right]\right\}.
\ee
Next, we estimate an average number of filaments $k$, associated with one observation pixel. We assume that the filaments are distributed randomly over a  device surface. Then for 52 non-volatile pixels we would have $52\times k$ filaments in all 7 samples. They are observed on the total surface of $7\times 81 \times 29\,\text{nm}^2$, which implies the area $S_0 \simeq 316/k\,\text{nm}^2$ per one filament. Hence, the probability to find $n$ filaments on a surface $S$ is quantified by the Poisson distribution, 
\bel{Poi}
P_n(S)= \frac{1}{n!}\left( \frac{S}{S_0}\right)^n e^{-S/S_0}.
\ee
Consider now the probability to have $k$ filaments on the surface, associated with the AFM tip, which is $s_0=29\,\text{nm}^2$. It reads $P_k(s_0)=(k!)^{-1}(s_0/S_0)^ke^{-s_0/S_0}= (k!)^{-1} (0.092\, k)^ke^{-0.092\,k}$, where we use $S_0\sim 316/k\,\text{nm}^2$. Hence we obtain $P_1=0.084$, $P_2=0.014$, $P_3=0.0026$, etc., which implies that only $k=1$ is consistent with our observations. Indeed, for $k=1$ we have $52\times 1 $ filaments for $81\times 7 =567$ pixels, so that the probability $P_1$ to find only one filament at any pixel is $52/567 \simeq 0.092$, very close to the above estimate of $P_1 = 0.084$. 

Similar analysis, performed for volatile filaments, indicates that the most probable is the case of $k=1$ filament per pixel. Therefore, we conclude that only one filament  contributes to the resistance of pixels -- both for volatile and non-volatile filaments. Then the resistance of the pixels can be estimated from Eq.~\eqref{Res2}. 

In Fig.~\ref{fig4} we plot the resistance of the pixel calculated from Eq.~\eqref{Res2} as a function of the energy gap $U_0-\e_F$ for two  different values of the distance between the last trap and bottom electrode, $l$. Here we take  $l_{nonvol}=1.1$\, nm  for non-volatile, and $l_{vol}=2.3$\, nm for volatile filaments. One can see that our estimate of resistance for both filament types is consistent with the experimental results and falls within the range of energy gap in $\alpha$-Si matrix, 2.00~eV - 2.75~eV.  The resistance of the non-volatile filaments, $l_{nonvol}=1.1$\, nm, for any bias voltage is stable and within the experimentally accessible range, while for the volatile filaments, $l_{vol}=2.3$\, nm, the resistance is measurable with a high bias voltage, $V$ = 7~V. At low bias voltage, $V$ = 0.5~V, resistance of volatile filaments is beyond the limits of our experimental setup.


We used conductive AFM to investigate the morphology and resistive switching in $\alpha$-Si/Ag/Cu memristors. Conductance in the device is a result of the formation of multiple and spatially distributed conductive filaments rather than uniform electron transport across the entire memristor area. Out of both volatile and non-volatile filaments, those showing volatile resistive switching dominate. The mean surface density of non-volatile filaments is about $3200$ filaments/$\mu\text{m}^2$. We believe that the leading transport in the filaments is a multiple trap assisted tunnelling, with the traps being $Ag^+$ ions (or the ions' clusters) in the bulk of $\alpha$-Si. The filament conductance is limited by the tunnelling between the last trap and the bottom electrode. The results demonstrate high stochasticity of the filament formation and conductivity, which is the main source of device-to-device and cycle-to-cycle variation in $\alpha$-Si/Ag/Cu memristors. By guiding the filament formation or confining them, one may suppress randomness and make resistive switching more uniform and reliable.

Authors acknowledge the support by the Russian Science Foundation (Grant No.~25-29-00544).


\bibliographystyle{apsrev4-1}
\bibliography{references}

\end{document}